\documentclass[a4paper,11pt]{scrartcl}

\pdfoutput=1

\makeatletter
\DeclareOldFontCommand{\rm}{\normalfont\rmfamily}{\mathrm}
\DeclareOldFontCommand{\sf}{\normalfont\sffamily}{\mathsf}
\DeclareOldFontCommand{\tt}{\normalfont\ttfamily}{\mathtt}
\DeclareOldFontCommand{\bf}{\normalfont\bfseries}{\mathbf}
\DeclareOldFontCommand{\it}{\normalfont\itshape}{\mathit}
\DeclareOldFontCommand{\sl}{\normalfont\slshape}{\@nomath\sl}
\DeclareOldFontCommand{\sc}{\normalfont\scshape}{\@nomath\sc}
\makeatother

 \usepackage{mathptmx}

\usepackage{hyperref}
\usepackage{pslatex}
\usepackage{amsmath}
\usepackage{txfonts}
\usepackage{graphicx}
\usepackage{cancel}
\usepackage{amssymb}
\usepackage{color}
\usepackage{authblk}
\graphicspath{{./}}

 \newcommand{\nn}{\nonumber}
 \newcommand{\NLO}{{\textrm{NLO}}}
 \newcommand{\NLOPS}{{\textrm{NLO+PS}}}

 \newcommand{\Refs}[1]{{refs.~\cite{#1}}}
 \newcommand{\Ref}[1]{{ref.~\cite{#1}}}
 \newcommand{\Eq}[1]{{eq.~(\ref{#1})}}
 \newcommand{\Fig}[1]{{fig.~\ref{#1}}}

\numberwithin{equation}{section}
\numberwithin{figure}{section}

\begin{document}
\titlehead{\hfill HU-EP-18/35}
\title{Matrix Element Method at NLO for (anti-)$\mathbf{k_t}$-jet algorithms}

 \author[]{Manfred Kraus\thanks{Manfred.Kraus@physik.hu-berlin.de} }
 \author[]{Till Martini\thanks{Till.Martini@physik.hu-berlin.de} }
 \author[]{Peter Uwer\thanks{Peter.Uwer@physik.hu-berlin.de}}
 \affil[]{\small Humboldt-Universit{\"a}t zu Berlin, Institut f{\"u}r Physik,
   Newtonstra{\ss}e 15, 12489 Berlin, Germany}

\maketitle

\begin{abstract}
  In this article, we present a method to calculate \textit{a posteriori} event
  weights at next-to-leading-order (NLO) QCD accuracy for a given jet event defined by the
  (anti-)$k_t$ algorithm relying on the conventional $2\to 1$
  recombination. This is an important extension compared to existing
  Monte-Carlo tools which generate jet events together with the
  corresponding weight but do not allow one to calculate the weight for a
  given event.  The method can be used to generate unweighted events
  distributed according to the fixed-order NLO cross section.  In
  addition, the method allows one to calculate NLO accurate weights for
  events recorded by experiments.  The potential of this ability is
  illustrated by applying the Matrix Element Method (MEM) to single
  top-quark events generated with \textsc{POWHEG} in combination with
  \textsc{Pythia}. For the first time, a systematic study of parton
  shower effects within the MEM is provided. The method is completely
  general and can be applied to arbitrary LHC processes. 
\end{abstract}

\newpage
\tableofcontents

\section{Introduction}
The steadily improving precision achieved in collider experiments like
ATLAS and CMS requires an equal precision in the theoretical
predictions to make optimal use of the experimental results. In recent
years tremendous progress has been made concerning the calculation of
next-to- and next-to-next-to-leading-order QCD corrections, see for instance~\Refs{Nason:2016tev,Caola:2015nfq,Heinrich:2017una,Duhr:2016nrb}. Meanwhile, next-to-leading-order corrections are
considered a solved problem and are
calculable for many processes using publicly available tools \cite{Badger:2012pg,Bevilacqua:2011xh,Cullen:2014yla,Alwall:2014hca,Gleisberg:2008ta,Campbell:2010ff,Frixione:2007vw}. As far as
next-to-next-to-leading-order corrections are concerned, the same
level of maturity has not been achieved yet.  However, many $2$-to-$2$
processes have been calculated recently \cite{Czakon:2016ckf,Boughezal:2013uia,Boughezal:2016wmq,Catani:2019iny,Bonciani:2018omm,Chen:2019zmr,Cruz-Martinez:2018rod,Currie:2017eqf,Ridder:2016nkl,Gehrmann:2014fva}.
  The forefront of current
research is the application to multiscale problems as they occur, for
example, in $2$-to-$3$ reactions \cite{Abreu:2019odu,Badger:2017jhb,Hartanto:2019uvl,Abreu:2018zmy} or $2$-to-$2$ processes involving particles
with many different masses \cite{Brucherseifer:2014ama,Caola:2017xuq,Grazzini:2018bsd}. For some quantities like, for example, the
Higgs cross section, even higher-order corrections have been calculated
recently \cite{Anastasiou:2015ema,Mistlberger:2018etf,Currie:2018fgr,Herzog:2017dtz}.

In collider experiments no smoking gun as a clear sign for physics
beyond the Standard Model has been observed so far. This has led to an
increasing interest in sophisticated analysis methods to compare
theoretical and experimental results (see e.g. \Refs{Brehmer:2019bvj,Bendavid:2018nar} and references therein). Multivariate methods allow one to utilise 
most of the information contained in the recorded events. 
Thus, they present promising tools to search for even smallest hints for New Physics. Among these
methods, the so-called Matrix Element Method sticks out since it
provides a very general and at the same time optimal approach to
compare theory and experiment. Based on the principle of Maximum
Likelihood, it allows for an unambiguous interpretation of the
findings. Briefly worded, the joint likelihood for a sample of
recorded events is calculated by interpreting the fully differential cross
section evaluated for each event as a measure for the 
probability of having measured this particular event. However,
until recently the use of the Matrix Element Method has been limited
by the fact that only leading-order cross sections could be used to
calculate weights for given events. Considering the progress in the
calculation of higher-order corrections mentioned above, this has been
a major drawback of the otherwise promising method.

Packages like \textsc{aMC@NLO} \cite{Alwall:2014hca} and \textsc{POWHEG} 
\cite{Frixione:2007vw} already allow the calculation of weights for
events beyond NLO accuracy. However, the ``direction'' of these
calculations is not compatible with the Matrix Element Method. In
\textsc{aMC@NLO} and \textsc{POWHEG}, one starts with a partonic momentum configuration
which is subsequently dressed with additional radiation. That approach
thus generates events with known event weights but does not allow one
to calculate the weight for a given event specified for example by
certain measured hadronic variables.  In contrast, the starting point
for the Matrix Element Method is a particular set of recorded events
with the corresponding weights to be calculated \textit{a posteriori}. To this
end the fully differential cross section is required as a function of
the event variables. In perturbation theory, the experimentally
resolved jets are modelled by mapping partonic momenta to jet momenta
according to the same jet algorithm that is used by the
experiment. The calculation of an event weight in perturbative QCD
therefore requires the cross section to be differential
in variables of the jet momenta modelling the recorded events. At
leading order, partonic and jet momenta are uniquely identified,
allowing for a straightforward evaluation of the event weight for
partonic final states.  However, when including higher-order
corrections, the jet algorithm dictates nontrivial mappings from
partonic momenta to jet momenta. Furthermore, this mapping depends on the phase space region. The 
identification of partonic momenta and jet momenta is thus no longer valid
beyond the leading order. The calculation of weights for measured
events in terms of cross sections which are differential in jet
variables is thus nontrivial when higher-order corrections are taken
into account.

 In \Refs{Alwall:2010cq,Campbell:2012cz}
different approaches to include NLO QCD corrections have been
investigated. However, both articles focus on special aspects and make
no attempt to present a general solution.  Reference~\cite{Alwall:2010cq}
concentrates on the effect of initial-state radiation, while 
\Ref{Campbell:2012cz} excludes strongly interacting particles in the
final state.
In \Ref{Martini:2015fsa} a general algorithm has been
proposed relying on a modification of the recombination procedure used
in jet algorithms. As a proof of concept, the method has been applied in
\Ref{Martini:2017ydu} to hadronic single top-quark production---a
process where the Matrix Element Method has been used from early days
on to disentangle the signal from an overwhelming
background \cite{Aad:2015upn}.

The evaluation of fully differential event weights incorporating NLO
QCD corrections also allows the generation of unweighted events
following the NLO differential cross sections. In fact, this
possibility has been used already in \Ref{Martini:2015fsa} and
\Ref{Martini:2017ydu} to simulate a toy experiment and to validate the
approach. These events obviously depend on the jet algorithm used to
cluster additional radiation. To distinguish them from partonic events
which are ill defined beyond leading order, we call them jet events in
the following. As outlined above, these jet events simulate---within
the fixed-order NLO approximation---the events observed in a real
experiment. It is well known that fixed-order predictions can be
further improved by resumming certain logarithmically enhanced
corrections through parton showers or analytic resummation. 
As mentioned before, packages like
\textsc{aMC@NLO} and \textsc{POWHEG} also allow the generation of unweighted events
including these effects. Although unweighted events following the
fixed-order NLO predictions may be considered as only halfway on the
way to unweighted events including parton-shower corrections, they are
interesting in their own right, since they allow a detailed study of
parton-shower effects. In addition, as has been pointed out in
\Ref{Figy:2018imt}, the possibility to produce unweighted jet events
may also help to improve the numerical integration over the real
corrections---a major bottleneck in the evaluation of NLO QCD cross
sections. 

At NLO QCD, real and virtual corrections are combined to calculate the
NLO contribution to differential cross sections. Because of additional
real radiation, the dimensionality of the phase space is different for
real and virtual corrections. As a consequence, the definition of a
``fully differential'' event weight requires integrating out all
unobserved radiation and adding this contribution to the virtual
corrections for the specific event. While this approach is
straightforward in theory, in practice, two complications arise:
\begin{enumerate}
 \item The variables used to describe the event must not allow one to distinguish
       between real and virtual contributions, since this would prevent the
       one-to-one correspondence of the two contributions. For example,
       in case of standard jet algorithms, relying on the summation of
       $4$-momenta to define the momentum of a jet obtained through
       recombination, the full $4$-momenta cannot be used to describe the event; the
       jets obtained from recombination will in general acquire a jet mass,
       and the pointwise correspondence of real and virtual corrections is
       lost. 
 \item For a given event, all additional radiation phase space
       which after recombination contributes to
       the event needs to be identified and integrated in an efficient way.
       While the combinatorial part for the possibilities to
       cluster the additional radiation is easy to
       solve, the efficient numerical integration is nontrivial. 
\end{enumerate}
Because of these two complications, the standard approach is to avoid the
definition of a fully differential event weight and combine finite
phase space regions in terms of histogrammed results as approximations
to differential distributions. However, drawbacks of this procedure
are potential numerical instabilities encountered at bin boundaries
and loss of information due to the binwise integration. For example,
unless high-dimensional histograms are used, correlations between
different event variables are lost.

The algorithm presented in \Refs{Martini:2015fsa,Martini:2017ydu} to
calculate weights for jet events at NLO accuracy modifies the
recombination procedure used to cluster two primary objects into a
resulting jet. It is thus possible to keep the kinematics Born-like,
leading to a straightforward identification of real and virtual
contributions without any further restrictions on the variables chosen
to describe the event. The second complication is solved by using in addition
the factorisation of the phase space in terms of a phase space for the
recombined jets and a part due to the additional radiation.  Although
the modification which clusters on-shell objects into on-shell jets is
theoretically well motivated, this recombination procedure is not yet
used in the experimental analysis since it would require a major
effort in recalibration and retuning of existing Monte Carlo
tools. To circumvent this problem, it has been shown in
\Ref{Martini:2018imv} for the example of single top-quark production
how the modification of the jet algorithm can be avoided, provided the
variables used to describe the event are carefully chosen. As
mentioned above, the basic idea is that the variables should not
allow one to reconstruct the invariant mass of the jets since outside soft
and collinear regions this precludes a one-to-one correspondence of
Born-like virtual corrections and contributions with additional real
radiation---which is required to uniquely define an event weight
incorporating NLO QCD corrections.  In fact, as we will show in the
next section, the method proposed in \Ref{Martini:2018imv} is rather
general and can be applied to arbitrary processes.

We note that similar ideas have been presented in
\Ref{Baumeister:2016maz,Figy:2018imt,Giele:2015sva,Weinzierl:2001ny}. In
\Ref{Baumeister:2016maz}, the problem is analysed from a mathematical
point of view, and a formal solution is given. However, the method
requires the numerical solution of a nonlinear system of equations
together with the numerical computation of the Jacobian for the
transformation. No proof of concept
that this can be done in a numerically stable and efficient way is given in \Ref{Baumeister:2016maz}. The
method presented here is very similar to the approach developed
in parallel in \Ref{Figy:2018imt}. The major difference is that in
\Ref{Figy:2018imt} an additional prescription to balance
the transverse momentum is used. In \Ref{Weinzierl:2001ny}, the generation 
of unweighted events of ``resolved'' pseudopartons is described, which can be 
used to calculate infrared observables at NLO accuracy.  

In fact, factorising the real phase space in terms of a Born-like phase 
space times the integration over the additional radiation is not a novel idea. 
In different contexts, it has been widely used to improve the efficiency of 
numerical phase space integrations \cite{Weinzierl:1999yf,Gleisberg:2008ta,Nagy:2003tz}. 

The article is organised as follows. In the next section, we describe
the method to calculate event weights including NLO QCD
corrections. 
In section~\ref{sec:example2}, the ability to predict event weights at NLO accuracy for jet events defined by 
conventional jet algorithms is employed in the Matrix Element Method. Exemplarily, events obtained from a
state-of-the art NLO+parton-shower event generator are analysed with the Matrix Element Method (MEM) at NLO. 
The top-quark mass extraction from single top-quark events is used to investigate the impact of parton 
shower effects on MEM-based analyses at NLO accuracy for the first time.
We emphasise that the extraction of the top-quark mass is only used to
have a concrete example. This work is not intended to strongly advocate the use of single
top-quark production to determine the top-quark mass. Nevertheless, the electroweak production of single top quarks does present a unique laboratory with the potential to study top-quark properties and compare them to results obtained from top-quark pair production (see, e.g., \Ref{Alekhin:2016jjz}).
A brief summary and the conclusions are given in the last section. 
Appendix~\ref{sec:example} summarises the concrete formulas implemented for the example application. Appendix~\ref{sec:valid} contains a validation of the calculated event weights. 

\section{NLO event weights for jet events defined by a $\bf{2\to1}$
  recombination procedure}\label{sec:evwgt21}

In this section, we describe the calculation of a fully differential
event weight including NLO QCD corrections. Fully differential means
in this context that the number of variables in which the cross
section is differential is maximal: $r=3n-4$ variables for the
production of $n$ jets in $e^+e^-$ annihilation and $r=3n-2$ variables
for the hadronic production of $n$ jets. We limit the discussion to
strongly interacting particles since the treatment of additional
particles which do not interact strongly is straightforward.  We
assume that an event for the process under consideration is described
by a set of $r$ variables $\{x_1,\ldots,x_r\}$. It is convenient,
although not necessary, to think of these variables as functions of the
momenta of the $n$ jets: $x_i = x_i(J_1,\ldots,J_n)$ or
$\vec{x} = \vec{ x}(J_1,\ldots,J_n)$ in vector notation, where the
$J_i$ are the $4$-momenta of the $n$ jets, which are calculated
according to the chosen jet algorithm.  Experiments usually record
values for variables related to energy depositions and particle tracks
in the detectors, which are attributed to jets. It is therefore also
natural to think of an event as a collection of experimentally
accessible variables $\{x_1,\ldots,x_r\}$ used to describe the
$4$-momenta of these resolved jets by imposing certain kinematics:
$J_1(\vec{x}),\ldots,J_n(\vec{x})$. 
We stress that using jets
automatically implies some ``inclusiveness'' even for fully differential
observables: to guarantee IR safety, the observable must be insensitive
to additional collinear and soft emission. In practice, this actually
means that soft and collinear regions of the real corrections are
integrated out. This can also be seen as some sort of averaging. Furthermore, it should be noted that in order to
calculate NLO-accurate corrections to the distribution of any variable
this variable must acquire nontrivial values already at the leading
order. Any variable which becomes nontrivial only at the NLO level
because of the presence of real corrections is only predicted with
leading-order accuracy. For example, the transverse
momentum of the $t\bar{t}$-system in top-quark pair production is only
leading-order accurate when the cross section for top-quark pair
production at next-to-leading order is used. To obtain NLO accuracy,
top-quark pair production in association with an additional jet needs
to be studied. Note that this feature is a general property of higher-order 
corrections. Restricting the variables to those present already
at the leading-order level automatically avoids the aforementioned problems.

The fully differential hadronic cross section including NLO corrections is then given by
\begin{eqnarray}\label{eq:master}
{d^r\sigma^{\NLO}\over dx_1\ldots dx_r}
  &=& \int dx_a dx_b\; d\Phi_n(x_a P_a +x_b P_b,\{p_1,\ldots,p_n\},
  \{m_1,\ldots,m_n\})\nn\\ 
  && \left[B+V\right](x_a,x_b;p_1,\ldots,p_n)\;
  \delta^{(r)}\left(\vec{x} - \vec{x}(J_1^{(n)}(p_1,\ldots,p_n),\ldots,
    J_n^{(n)}(p_1,\ldots,p_n)\right)\nn\\
  &+& \int dx_a dx_b\; d\Phi_{n+1}(x_a P_a +x_b P_b,
  \{p_1,\ldots,p_{n+1}\},\{m_1,\ldots,m_{n+1}\})\nn\\
  &&R(x_a,x_b;p_1,\ldots,p_{n+1})\;
  \delta^{(r)}\left(\vec{x} - \vec{
      x}(J_1^{(n+1)}(p_1,\ldots,p_{n+1}),\ldots,
    J^{(n+1)}_n(p_1,\ldots,p_{n+1})\right)\;,\nn\\
\end{eqnarray}
where $[B+V]$ denotes the sum of the Born and virtual contributions
and $R$ denotes the real corrections due to additional radiation. 
Real and virtual contributions 
are in general individually IR divergent. In what follows, a suitable
prescription to handle these singularities is always implicitly
understood. The case of $e^+e^-$-annihilation is
straightforward once hadronic collisions are understood; therefore, we limit our 
discussion to the latter ones.
To simplify the notation we have absorbed the parton
distribution functions into the functions $[B+V]$ and $R$. The momenta
of the incoming hadrons $a$ and $b$ are given by $P_a$ and $P_b$ and
the parton-momentum fractions are denoted by $x_a$ and $x_b$. 
The functions $J_i^{n}$ and $J_i^{(n+1)}$ encode the construction of $n$ jet momenta from either $n$ or $n+1$ partonic momenta according to the jet algorithm defined by the resolution criterion and the recombination procedure.   
The phase
space measure is given by
\begin{equation}
  d\Phi_n(P,\{p_1,\ldots,p_n\},\{m_1,\ldots,m_n\}) = (2\pi)^4 
  \delta\left(P-\sum_{k=1}^n p_k\right) 
  \prod_{k=1}^n {d^4p_k\over (2\pi)^3} \delta\left({p_k}^2-{m_k}^2\right)\;. 
\end{equation}
To allow an efficient numerical evaluation, the delta-functions in
\Eq{eq:master} need to
be integrated out analytically. In the Born and virtual corrections, this
corresponds to the solution of a system of equations implied by
the delta-functions, since all integrations are fixed through the
delta-functions. Another way to phrase this is to rewrite the phase space 
$d\Phi_n$ in terms of the variables  $\{x_1,\ldots,x_r\}$,
\begin{equation}\label{eq:Bornpsfacx}
  d\Phi_n(x_a P_a +x_b P_b,\{p_1,\ldots,p_n\},\{m_1,\ldots,m_n\}) =
  \hat{\cal J}(\vec{x})\; dx_1\ldots dx_r\;,
\end{equation}
with Jacobian $\hat{\cal J}(\vec{x})$, which allows for a trivial integration over the delta-functions. 
In case of the real corrections, a similar factorisation of the form 
\begin{eqnarray}
  d\Phi_{n+1}(x_a P_a +x_b P_b,\{p_1,\ldots,p_{n+1}\},\{m_1,\ldots,m_{n+1}\})\Big|_{{i\text{ unres.}}} 
  =\tilde{{\cal J}}(\vec{x},p_i)\; 
  dx_1\ldots dx_r\; d\Phi_i
\end{eqnarray}
is required, where $d\Phi_i$ denotes the phase space associated with
the ``unresolved'' radiation of parton $i$ and its respective Jacobian $\tilde{{\cal J}}(\vec{x},p_i)$.
In general, for arbitrary variables $\{x_1,\ldots,x_r\}$, such
a factorisation does not exist because the jets $J_k^{(n+1)}$
obtained from the recombination of two partons do not necessarily
satisfy the same kinematical constraints as the ones used in the
virtual corrections, e.g.,
\begin{eqnarray}\label{eq:onshellmomcons}
  (J_k^{(n+1)})^2 \not= (J_k^{(n)})^2={p_k}^2={m_k}^2,\qquad 
  \sum\limits_k {\overrightarrow{J_k^{(n+1)}}}^{\perp}\not= \sum\limits_k {\overrightarrow{J_k^{(n)}}}^{\perp}=0.
\end{eqnarray}
Note that in the Born and virtual contributions no recombination takes place.
In this article, we propose a solution while employing the commonly used recombination, where 
the $4$-momentum of a jet obtained from the clustering of two partons $i,j$ is given by the sum
of the momenta $p_i$ and $p_j$~\cite{Cacciari:2011ma},
\begin{equation}\label{eq:escheme}
  J_{(ij)}\equiv J^{(n+1)}_{(ij)}(p_1,\ldots,p_i,\ldots,p_j,\ldots,p_{n+1}) 
  = p_i+p_j
\end{equation}
with
\begin{equation}\label{eq:jetmass}
 (J_{(ij)})^2=(p_i+p_j)^2\;.
\end{equation}
The remaining $n-1$ jet momenta are identified with the partonic
momenta. Because the jet masses
$(J_{(ij)})^2$ are in general different from the ones occurring in the
virtual corrections, the variables $\{x_1,\ldots,x_r\}$ must not depend
on the jet masses, since this would prevent the pointwise combination
of real and virtual contributions. In fact, this requirement can also
be understood as a consequence of infrared safety. 

Due to the different parton multiplicities of the phase space
integrations in \Eq{eq:master}, it is convenient to split the cross
section into the
Born and virtual contribution ($\text{BV}$) and the real contribution
($\text{R}$). The real phase space can be further partitioned into
regions ${\cal R}_{ij}$ where the parton pair $ij$ to be
clustered is picked  by the respective jet algorithm,
regions ${\cal R}_{i}$ where one parton $i$ escapes detection, and
regions ${\cal \tilde R}_{i}$ where all partons are resolved as jets
but (the softest) jet $i$ is not considered in the event definition:
\begin{equation}\label{eq:masterfac}
 {d^r\sigma^{\NLO}\over dx_1\ldots dx_r}=
  {d^r\sigma^{\text{BV}}\over dx_1\ldots dx_r}
  \hspace{1ex}+\hspace{1ex} 
  \sum_{i}\sum_{\substack{j\\j\neq i}}  {d^r\sigma^{\text{R}}_{{{\cal R}_{ij}}}\over dx_1\ldots dx_r}
   \hspace{1ex}+ \hspace{1ex}\sum_{i}
 {d^r\sigma^{\text{R}}_{{{\cal R}_{i}}}\over dx_1\ldots dx_r}
  \hspace{1ex}+\hspace{1ex} \sum_{i}{d^r\sigma^{\text{R}}_{{\tilde{\cal
          R}_{i}}}\over dx_1\ldots dx_r
    }.
\end{equation}
The sums run over all $i,j$, which can be clustered/omitted to still end up 
with the signal signature of the Born process.
The last term is absent if additional jet activity is vetoed.  The
differential cross section can be interpreted as an event weight at
NLO accuracy for the event defined by the variables $\{x_1,\ldots,x_r\}$.
The technical implementation of the phase space partitioning is explained in detail in 
\Refs{Martini:2015fsa,Martini:2017ydu,Martini:2018imv} and exemplified in Appendix~\ref{sec:example} .

In the following, we give explicit parametrisations of the $n$- and
$(n+1)$-parton phase space, allowing for the combination of virtual and
real corrections.  Since we have to show that it is possible to
factorise $d\Phi_{n+1}$ into a Born-like phase space times some
additional contribution due to extra radiation, we start with
$d\Phi_n$.  In the Born and virtual part, the $n$ parton momenta are
identified with the momenta of the $n$ jets. A useful, although not
unique, starting point is given by the following parametrisation (cf.
\Eq{eq:Bornpsfacx}),
\begin{eqnarray}\label{eq:bornps}
  && dx_adx_b\;d\Phi_n(x_a P_a+x_bP_b,
  \{p_1,\ldots,p_n\},\{m_1,\ldots,m_n\})\nn \\
  &=& \;{(2\pi)^{(4-3n)}\over 2^{n-1} } {1\over s}
  \;{\cal J}(\vec{x})\;dx_1\ldots dx_r\nn\\
  \nn && \times\;  \left[\prod_{k=1}^n d{J_k}^2\; 
    \delta\left({J_k}^2-{m_k}^2\right){1\over J_k^0}\right]\;
  d^2J^{\perp}_1\;\delta^{(2)}
  \left(\vec{J}^{\perp}_1+\sum_{{k=2}}^n 
    \vec{J}^{\perp}_k\right)\nn\\
  &&\times\; dx_a dx_b\;
  \delta\left(x_a-{1\over \sqrt{s}} \sum_{k=1}^n (J^0_k+J^z_k)\right) \;
  \delta\left(x_b-{1\over \sqrt{s}}\sum_{k=1}^n (J^0_k-J^z_k)\right)\; 
  ,
\end{eqnarray}
where $s=(P_a+P_b)^2$ and ${\cal J}(\vec{x})$ is the Jacobian of the variable transformation 
\begin{equation}\label{eq:vartrf}
\left\{J_1^z,\vec{J}_{2},\ldots,\vec{J}_{n}\right\}\mapsto \{x_1,\ldots,x_r\}\quad\text{ with }\quad{\cal J}(\vec{x})=
  \left|\partial\left(J^z_1,J^x_2,J^y_2,J^z_2,\ldots,J^x_n,J^y_n,J^z_n\right)\over
  \partial(x_1,\ldots, x_r)\right|.
\end{equation}
If the events are defined by the longitudinal component of one of the resolved jets 
and the $n-1$ $3$-momenta of the remaining ones, then this is just an identity transformation 
with ${\cal J}(\vec{x})=1$. In Appendix~\ref{sec:example}, the explicit transformation and respective 
Jacobian is given for an event definition in terms of energies and angular variables for the case of single
top-quark production.

The jet masses, the transverse momentum, and the variables
$\{x_1,\ldots,x_r\}$ are used as integration
variables. On-shell conditions and momentum conservation fix the
former, leaving the remaining variables $\{x_1,\ldots,x_r\}$ to 
define a jet event.  This factorisation thereby allows a
straightforward integration of the delta-function
\begin{displaymath}
\delta^{(r)}\left(\vec{x}-\vec{x}(J^{(n)}_1\equiv J_1,\ldots,J^{(n)}_n\equiv J_n)\right)
\end{displaymath}
in the Born and virtual contributions in
\Eq{eq:master}. With this parametrisation, the Born and virtual part of
the differential cross section reads
\begin{equation}\label{eq:dxsbvpart}
  {d^r\sigma^{\text{BV}}\over d{x}_1\ldots d{x}_r}
  ={\cal J}(\vec{x})\;{(2\pi)^{(4-3n)}\over 2^{n-1} }{1\over s}\; 
  \left[\prod_{k=1}^n{1\over J_k^0}\right]
  \left[B+V\right]\left(x_a,x_b;p_1(\vec{x}),\ldots,p_n(\vec{x})\right)\;.
\end{equation}
 The parametrisation of the $n$ partonic momenta $\{p_1,\ldots,p_n\}$ in terms of 
$\{x_1,\ldots,x_r\}$ is fixed by the delta-functions in
\Eq{eq:bornps} and the identification of $n$ partonic with $n$ jet momenta.

Let us now study the real contribution. We focus on those regions of
the ($n+1$) parton phase space in which additional radiation is
recombined into a jet or associated with the beam, 
resulting in an $n$-jet final state. The two cases need to be treated
separately because the underlying phase space factorisation is inherently different. The
remaining regions where the additional radiation is resolved as an
additional jet do not impose any conceptual problems and can be
obtained from a similar parametrisation of the real phase space as
presented below.

First, we consider the situation in which the additional radiation is
clustered with a final-state parton. We denote with ${\cal R}_{ij}$ a
phase space region in which the two final-state partons $i$ and $j$
are combined according to \Eq{eq:escheme} to form a jet $(ij)$ with invariant
mass squared,
\begin{displaymath}
  {M_{ij}}^2 \equiv (J_{(ij)})^2={(p_i+p_j)^2}.
\end{displaymath}
Note that the momenta of the remaining jets are identified with the
underlying partonic momenta as in the Born and virtual contributions.
\begin{figure}[htbp]
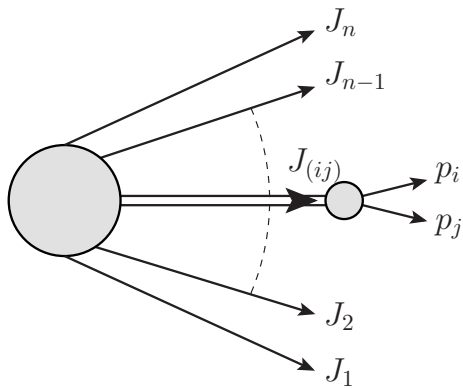

  \begin{center}
    \leavevmode
   \includegraphics[width=0.49\textwidth]{{{%
         21clusn}}}
    \caption{Factorisation of the real phase space for the 
      $2\to1$ clustering of partons $i,j$ into an off-shell Born and a two-particle phase space.}
    \label{fig:21psfac}
  \end{center}
\end{figure} 
By utilising a factorisation of the real phase space
corresponding to the clustering given in \Eq{eq:escheme}, we obtain
(cf. \Fig{fig:21psfac})
\begin{eqnarray}
  \label{eq:realpsfac}
  &&
  d\Phi_{n+1}(P,\{p_1,\ldots,p_i,\ldots,p_j,\ldots p_{n+1}\},
  \{m_1,\ldots,m_i,\ldots,m_j,\ldots m_{n+1}\})\nn\\
  &=& 
  d\Phi_n(P,\{J_1,\ldots,\cancel{p_i},\;J_{(ij)},\ldots,\cancel{p_j},\ldots, J_{n}\},
  \{m_1,\ldots,\cancel{m_i},\;M_{ij},\ldots,\cancel{m_j},\ldots, 
  m_{n+1}\})\nn\\
  &\times& (2\pi)^{-1} d{M_{ij}}^2\;
  d\Phi_2(J_{(ij)},\{p_i,p_j\},\{m_i,m_j\})\;.
\end{eqnarray}
Using the parametrisation from the Born and virtual part for the $n$-particle phase space (with the off-shell momentum $J_{(ij)}$),
\begin{displaymath}
{d\Phi_n(P,\{J_1,\ldots,\cancel{p_i},\;J_{(ij)},\ldots,\cancel{p_j},\ldots, J_{n}\},
  \{m_1,\ldots,\cancel{m_i},\;M_{ij},\ldots,\cancel{m_j},\ldots, 
  m_{n+1}\})}
\end{displaymath}
the same $(3n-2$)
variables $\{x_1,\ldots,x_r\}$ as in \Eq{eq:bornps} appear as
independent variables. It is thus straightforward to integrate out the
second delta-function
\begin{displaymath}
\delta^{(r)}\left(\vec{x}-\vec{x}(J^{(n+1)}_1\equiv J_1,\ldots,J^{(n+1)}_n\equiv J_n)\right)
\end{displaymath}
in \Eq{eq:master} while the unresolved parton pair is to be integrated over the region ${\cal R}_{ij}$.  The contribution of the real
corrections to the differential cross section 
from the region ${\cal R}_{ij}$ reads
\begin{eqnarray}\label{eq:dxs21part}
  {d^r\sigma^{\text{R}}_{{{\cal R}_{ij}}}\over d{x}_1\ldots d{x}_r }
 &=&{(2\pi)^{(3-3n)}\over2^{n-1}}{1\over s}\; 
 \int_{{{\cal R}_{ij}}}{d{M_{ij}}^2}\;{\cal J}(\vec{x})\; \left[\prod_{{k=1}}^{n}{1\over
     J^0_{k}}\right]\;
 d\Phi_2(J_{(ij)},\{p_i,p_j\},\{m_i,m_j\})\nn\\
 &\times&  R\left(x_a,x_b;p_1(\vec{x},{M_{ij}},\phi_i,\theta_i),\ldots,p_{n+1}(\vec{x},{M_{ij}},\phi_i,\theta_i)\right)\;. 
\end{eqnarray}
The parametrisation of the $n+1$ partonic momenta
$\{p_1,\ldots,p_{n+1}\}$ in terms of $\{x_1,\ldots,x_r\}$, ${M_{ij}}$
and the azimuthal and polar angle $\phi_i$ and $\theta_i$ of parton
$i$ is fixed by the delta-functions in \Eq{eq:bornps} and the
clustering of $n+1$ partonic momenta to $n$ jet momenta in the region
${\cal R}_{ij}$. The Jacobian  ${\cal J}(\vec{x})$ occurring in \Eq{eq:dxs21part}
is the same as in \Eq{eq:dxsbvpart}.

Let us now investigate the contribution from regions ${\cal R}_i$
in which the parton $i$ is considered as part of the beam and needs to
be integrated out. Since there is no recombination in this case, the
momenta of the resolved jets are identified with the underlying
partonic momenta.  
\begin{figure}[htbp]
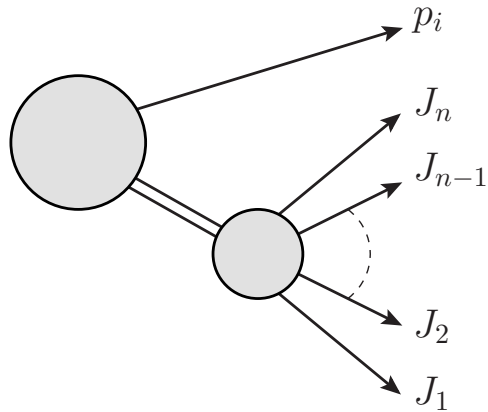

  \begin{center}
    \leavevmode
  \includegraphics[width=0.49\textwidth]{{{%
         21clusni}}}
    \caption{Real phase space for the 
     unresolved parton $i$ as boosted Born and additional parton.}
    \label{fig:21psifac}
  \end{center}
\end{figure} 
The real phase space $d\Phi_{n+1}$ can be factorised in terms of a boosted $n$-particle phase space 
and integrations over the degrees of freedom of the additional parton (cf. \Fig{fig:21psifac}). 
The momenta in the boosted $n$-particle phase space, corresponding to the $n$ resolved jets, can 
be parameterised by the same $(3n-2)$ variables as in \Eq{eq:bornps}:
\begin{eqnarray}\label{eq:realise}
  \nn &&
  dx_adx_b\;d\Phi_{n+1}(x_aP_a+x_bP_b,\{p_1,\ldots,p_{n+1}\},
  \{m_1,\ldots,m_{n+1}\})\\
  \nn&=&{d^4p_i \over (2\pi)^3}\delta\left({p_i}^2-{m_i}^2\right)\;dx_adx_b\;d\Phi_{n}(x_aP_a+x_bP_b-p_i,\{J_1,\ldots,J_{n}\},
  \{m_1,\ldots,m_{n}\})\\
  \nn&=&  {(2\pi)^{(4-3n)}\over2^{n-1}} {1\over s}
  \;{\cal J}(\vec{x})\;dx_1\ldots dx_r\;{d^3p_{i}}\\
  &&\times\;{(2\pi)^{-3}\over 2}\left[\prod_{k=1}^{n} 
    d{J_k}^2\; \delta\left({J_k}^2-{m_k}^2\right)\;{1\over J_k^0}\right]\; 
  {d{p_i}^2\over p_i^0}\; \delta\left({p_i}^2-{m_i}^2\right)\;
  d^2J^{\perp}_1\;
  \delta^{(2)}\left(-\vec{p}^{\perp}_i-\sum_{{k=1}}^{n} \vec{J}^{\perp}_k\right)\nn\\
  && \times\; dx_a  dx_b\; \delta\left(x_a-{ p^0_{i}+p^z_{i}\over \sqrt{s}}
   -{1\over \sqrt{s}}
    \left(\sum_{k=1}^{n} 
      (J^0_k+J^z_k)\right)\right) 
  \;\delta\left(x_b-{p^0_{i}-p^z_{i}\over \sqrt{s}}-{1\over \sqrt{s}}\left(\sum_{k=1}^{n} 
      (J^0_k-J^z_k)\right)\right)\;. \nn\\
\end{eqnarray}
As in the previous case, the jet masses, the transverse momentum, and the $3$-momentum of the
unresolved radiation appear as additional integration variables. 
On-shell conditions and momentum conservation again fix the former,
while the $3$-momentum $\vec{p}_i$ and the variables
$\{x_1,\ldots,x_r\}$ are left as integration variables. It
is thus straightforward to integrate out the second delta-function
\begin{displaymath}
\delta^{(r)}\left(\vec{x}-\vec{x}(J^{(n+1)}_1\equiv J_1,\ldots,J^{(n+1)}_n\equiv J_n)\right)
\end{displaymath}
in \Eq{eq:master}, leaving the unresolved radiation to be integrated over the region ${\cal R}_{i}$. 

The contribution of the real corrections from the region 
${\cal R}_{i}$ to the differential cross section is then given by
\begin{eqnarray}\label{eq:dxsinipart}
  {d^r\sigma^{\text{R}}_{{{\cal R}_{i}}}\over d{x}_1\ldots d{x}_r}
  &=&{(2\pi)^{(1-3n)}\over2^{n}}{1\over s}\;
   \int_{{{\cal R}_{i}}}{d^3p_{i}\over p^0_{i}}\;{\cal J}(\vec{x})\;   \left[\prod_{k=1}^n{1\over J_k^0}\right]\;
  R\left(x_a,x_b;p_1(\vec{x},\vec{p}_i),\ldots,p_{n+1}(\vec{x},\vec{p}_i)\right)\;,\nn\\
\end{eqnarray}
where the parametrisation of the $n+1$ partonic momenta $\{p_1,\ldots,p_{n+1}\}$ in terms of
$\{x_1,\ldots,x_r\}$ and $\vec{p}_i$ is fixed by the delta-functions in \Eq{eq:realise} and 
the identification of $n$ resolved parton momenta with $n$ jet momenta.

It is worth mentioning that with a slight modification of \Eq{eq:dxsinipart}
the contribution from the real corrections where the extra radiation
$i$ is resolved as an additional jet with momentum $J_i$ but does
not enter the event definition can be obtained: the
integration over $\vec{J}_i$ has to be carried out over the region
$\tilde {\cal R}_i$ of the real phase space where parton $i$ is
resolved as an additional jet,
\begin{eqnarray}\label{eq:dxsaddjpart}
  {d^r\sigma^{\text{R}}_{\tilde{\cal R}_i}\over d{x}_1\ldots d{x}_r}
  = \int_{\tilde{\cal R}_i}{d^3J_{i}\over J^0_{i}}\;{\cal J}(\vec{x})\;   \left[\prod_{k=1}^n{1\over J_k^0}\right]\;
  R\left(x_a,x_b;p_1(\vec{x},\vec{J}_i),\ldots,p_{n+1}(\vec{x},\vec{J}_i)\right)\;.\nn\\
\end{eqnarray}
Collecting the Born and virtual contributions (see \Eq{eq:dxsbvpart})
and the real corrections from the different regions ${\cal
  R}_{ij}$ (see \Eq{eq:dxs21part}), ${\cal R}_i$ (see
\Eq{eq:dxsinipart}), and $\tilde {\cal R}_i$ (see \Eq{eq:dxsaddjpart}),
we finally obtain the fully differential cross section including the 
NLO corrections (cf. \Eq{eq:masterfac}).
A validation of this approach for the example of single top-quark production is given in Appendix~\ref{sec:valid}.

\section{Parton shower effects in fixed-order MEM analyses}\label{sec:example2}
In previous applications of the
MEM at NLO to fully hadronic processes, the
approach has been restricted by requiring a $3 \to 2$ clustering in the jet
algorithm~\cite{Martini:2015fsa,Martini:2017ydu}.  Therefore, only
closure tests of the method, where
the pseudodata have been generated with the modified clustering
prescription, have been presented so far. The method
presented in this paper and \Ref{Martini:2018imv} extends the MEM at
NLO to incorporate $2\to 1$ clusterings and therefore allows one to
analyse realistic events.

We use the \textsc{POWHEG-BOX}~\cite{Alioli:2009je,Alioli:2010xd}
to generate pseudodata.  The events are subsequently showered using
the \textsc{Pythia8} parton-shower (PS) program~\cite{Sjostrand:2014zea}.
During the parton-shower evolution, the top quark is kept stable, and
corrections from hadronisation and underlying events have been
neglected. Thus, the main difference between the pseudodata and the
fixed-order NLO calculation used for the analysis is the inclusion of
additional radiation in the events. This allows us for the first time to study
the impact of parton-shower effects within the MEM at NLO accuracy.

For a concrete example, we study the top-quark mass determination from hadronic single 
top-quark production via the $t$-channel:
\begin{equation}\label{eq:sigsig}
pp\rightarrow tj\;.
\end{equation}
The event signature is given by a top-tagged jet\footnote{A top-tagged jet is a jet containing a top quark.} $t$
in association with at least one light jet $j$.  The sample is generated
for the LHC at $\sqrt{s} = 13$ TeV and a top-quark mass of $m_t =
173.2$ GeV. The renormalisation and factorisation scales are set to $\mu_R =
\mu_F = \mu_0 = m_t$.  Jets are defined using the
$k_t$-algorithm \cite{Cacciari:2011ma}  with a separation parameter of $R = 0.4$.
Unresolved partons are clustered by the $2\to1$ recombination by summing their $4$-momenta.
Resolved jets have to pass the following cuts on the
transverse momentum and the pseudorapidity
\begin{equation*}\label{eq:expcuts}
 p^\perp > p^\perp_{\text{min}}=30\text{ GeV},\qquad\quad |\eta|< \eta_{\text{max}}= 3.5.
\end{equation*}  
The detector is assumed to be blind outside these cuts.
If there is more than one resolved light jet, the
hardest one is used in the event definition. We
assume that it is always possible to identify the jet containing the
top quark.
The resulting event sample for the fiducial phase space volume consists of $N=28031$
events.  

The top-quark mass extraction proceeds as follows. For each showered
\textsc{POWHEG }event $i$, the set of variables
$({\eta_t}^i,{E_j}^i,{\eta_j}^i,{\phi_j}^i)$ is calculated from the
top-tagged jet and the hardest light jet. The motivation for this
choice of variables is two-fold. First of all, measuring angular
variables is experimentally under better control than variables
related to the energy of final-state objects. This is especially true
for jets, since the jet-energy scale often represents a major
uncertainty. By preferring the former over the latter, the deviations
from trivial transfer functions should only be a moderate
effect. Second, this set of variables is equivalent to the generic
set given in \Eq{eq:vartrf} with the respective variable
transformation given in Appendix~\ref{sec:example}. As long as the
Jacobian is independent of the parameter to be estimated, the MEM and
its results are invariant under any change of variables complying with
the requirements of not fixing the jets' masses and the overall
transverse momentum given in section~\ref{sec:evwgt21}. However, it
should be stressed that not all variables can be reached from the
aforementioned ones by a variable transformation. This is not a
restriction \textit{per se}. It simply means that because of the intrinsic
inclusiveness of the jets the contribution of the real corrections---the
unresolved regions which need to be integrated out---depends on the
chosen variables. For variables not reachable  by a variable
transformation, one can still use the procedure presented in this
article. It would just mean that one has to start with a different set
of variables.

\begin{figure*}[h!]
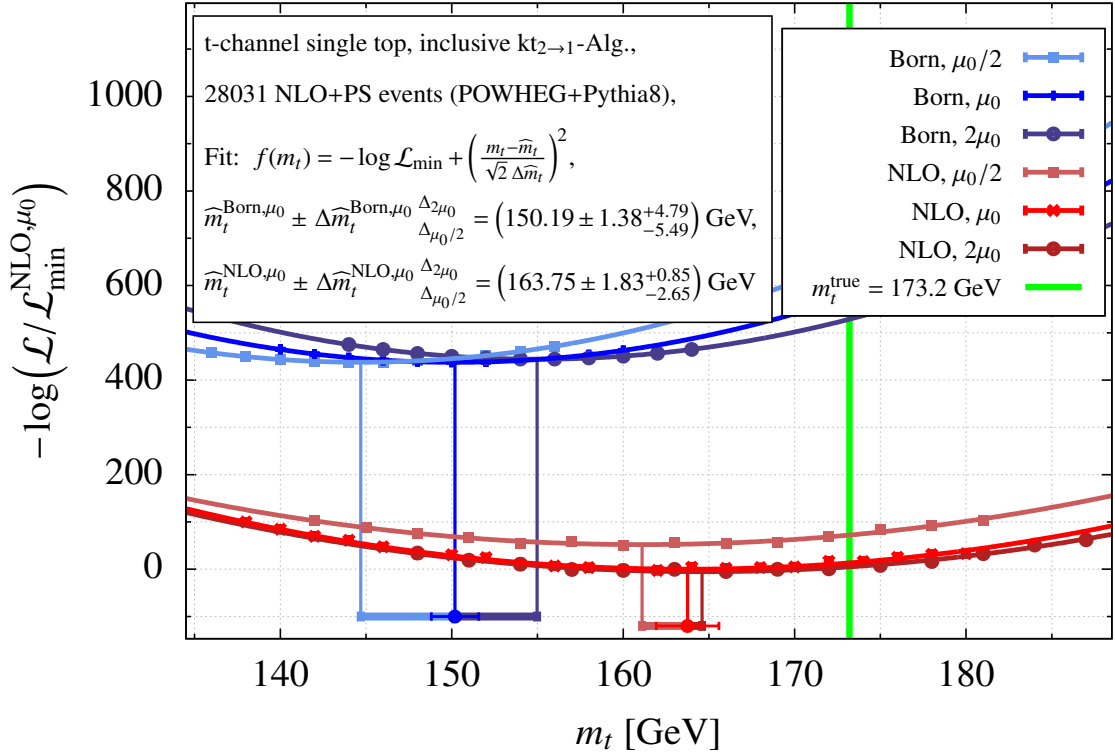

  \begin{center}
    \leavevmode
     \includegraphics[width=0.99\textwidth]{{{%
          sgttKT2-1ycut30memtwobornnloincmu-113evts-crop}}}
    \caption{Top-quark mass extraction with the Matrix Element Method from $t$-channel single top-quark events at the LHC generated with \textsc{POWHEG} and \textsc{Pythia8}.}
    \label{fig:MEM_sgtt21}
  \end{center}
\end{figure*}

For each event, the weight is calculated 
for a specific value of the top-quark mass (cf. \Eq{eq:masterfac} 
and details in Appendix~\ref{sec:example}). The corresponding likelihood 
function
$\mathcal{L}(m_t)$ is computed according to
\begin{equation}
 \mathcal{L}(m_t) = \prod_{i=1}^N \frac{1}{\sigma(m_t)}\frac{d^4\sigma(m_t)}{d{\eta_t}^i\;d{E_j}^i\;d{\eta_j}^i\;d{\phi_j}^i}
 ,
 \label{eqn:Likelihood}
\end{equation}
where only trivial transfer functions (i.e., delta-functions) have been
used.  For more details, we
refer to \Ref{Martini:2018imv}. Minimizing the negative
log-likelihood function $(-\log\mathcal{L})$ with respect to the
top-quark mass parameter yields the estimator for the top-quark mass
$\widehat{m}_t$. We have checked that we correctly reproduce the input
value if the pseudodata are generated using fixed-order predictions and
parton-shower effects are neglected.  Figure~\ref{fig:MEM_sgtt21} shows
the negative log-likelihood function for the \textsc{POWHEG} events as a function of the top-quark
mass and for various scale choices at leading order (LO) and NLO in perturbation
theory.  The leading-order results are shown in blue for three
different scales, the central scale $\mu_0$ (blue) and scale
variations $\mu_0/2$ (light blue) as well as $2\mu_0$ (dark blue).
The red curves are obtained by including the NLO QCD corrections in
the definition of the likelihood functions and are also shown for
three different scales: $\mu_0$ (red), $\mu_0/2$ (light red), and $2\mu_0$
(dark red).  The vertical green line denotes the value for the
top-quark mass used for the generation of the pseudodata. The minimum
of the curves yields the estimator for the top-quark mass
$\widehat{m}_t$ and the width of the parabola gives the statistical
uncertainty $\Delta\widehat{m}_t$ on the estimator.  The estimated
top-quark mass using the Born approximation for the central scale is
thus given by $\widehat{m}_t = (150.19 \pm 1.38_{\text{stat}})$
GeV. Based on the results for the two other scale settings, the
uncertainty due to neglected higher-order corrections is estimated to
be of the order of $\pm 5$ GeV. We observe a significant difference
between the results obtained from the MEM using LO matrix elements and
the input value. In particular, the observed shift of about 22 GeV is
not covered by the uncertainties. Using the MEM based on leading-order
matrix elements thus requires a significant calibration to reproduce
the input value.

Using, on the other hand, the full NLO QCD calculation for the determination of the event
weights entering the likelihood function, the estimator for the
top-quark mass gives $\widehat{m}^\NLO_t = (163.75 \pm 1.83_{\text{stat}})$ GeV. A
significant reduction of the scale uncertainty is observed.  The
uncertainty goes down from the aforementioned $\pm 5$ GeV to $+1$
GeV and $-3$ GeV at NLO. Furthermore the shift compared to the input value
is reduced to 10 GeV compared to 22 GeV at leading order.

In fact, a shift in the extracted top-quark mass to lower mass values
is expected due to parton-shower effects in the pseudodata which are
not taken into account in the MEM: multiple parton emissions lead to a
modification of the phase space density, which results in shape differences in
differential distributions compared to fixed-order NLO computations.
The MEM is very sensitive to small distortions
of the differential distributions.  As the parton-shower tends to
soften the $p_{\perp}$ distribution, the MEM favours a smaller mass value.

\begin{figure*}[h!]
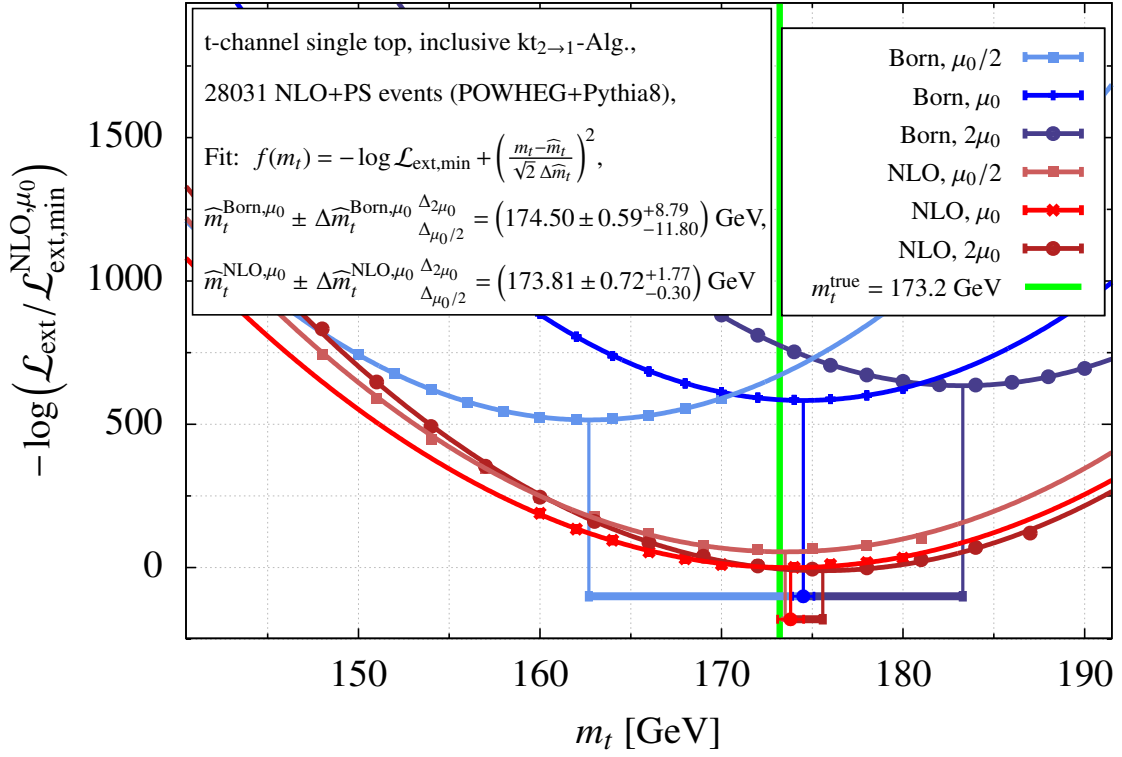

  \begin{center}
    \leavevmode
     \includegraphics[width=0.99\textwidth]{{{%
          sgttKT2-1ycut30memtwobornnloincmu-113evts_ext-crop}}}
    \caption{Top-quark mass extraction with the Extended Matrix
      Element Method from $t$-channel single top-quark events at the
      LHC generated with \textsc{POWHEG} and \textsc{Pythia8}.}
    \label{fig:MEM_ext_sgtt21}
  \end{center}
\end{figure*}

Note that the analysis based on the likelihood function given in
\Eq{eqn:Likelihood} is only sensitive to the normalised
multidifferential cross section.  As argued in
\Ref{Martini:2017ydu,Martini:2018imv}, the information about the
fiducial cross section can improve the parameter determination. To
incorporate the information on the total number of events in the sample, the extended
likelihood function, defined by
\begin{equation}
  \mathcal{L}_{\text{ext}}(m_t) = {{\nu(m_t)^N}\over{N!}} e^{-\nu(m_t)} \mathcal{L}(m_t) =\frac{L^N}{N!}e^{-\sigma(m_t)L} \prod_{i=1}^N \frac{d^4\sigma(m_t)}{d{\eta_t}^i\;d{E_j}^i\;d{\eta_j}^i\;d{\phi_j}^i}
 ,
  \label{eqn:extLikelihood}
\end{equation}
is used. Here, $\nu(m_t) = L\cdot \sigma(m_t)$ is the predicted number of events, where $L$ denotes 
the integrated luminosity of the
experiment.  For the pseudodata used in the analysis, the integrated luminosity is given by
$N = L\cdot\sigma^\NLOPS$, where $\sigma^\NLOPS$ is the fiducial cross
section corresponding to the simulated data. 
Including the information on the number of recorded events should give a
significant improvement: while the parton shower tends to soften the
$p_{\perp}$ distribution leading to the aforementioned smaller mass values, the
total number of events is only mildly affected through acceptance effects.
Since a smaller mass value leads to larger cross sections and thus a
larger number of events, including this information through the extended likelihood
increases the sensitivity of the analysis.
The results of the extended likelihood analysis are
shown in Fig.~\ref{fig:MEM_ext_sgtt21}.
The fit improves significantly compared to the previous analysis. At
LO and NLO accuracy, the estimated mass value is now compatible with
the input value.  Including the information about the normalisation,
the Born approximation yields a top-quark mass estimator of
$\widehat{m}_t = 174.50$ GeV with a statistical uncertainty of
$\Delta\widehat{m}_t =\pm 0.59$ GeV. While the statistical uncertainty
is reduced by more than a factor of $2$, the systematic uncertainty due
to scale variations is enlarged by more than a factor of $2$, which
amounts to roughly $\pm 10$ GeV.  Repeating the extraction including
the NLO QCD corrections in the extended likelihood function improves
the analysis significantly with respect to the Born approximation. The
top-quark mass estimator and its statistical uncertainty are given by
$\widehat{m}^\NLO_t = (173.81 \pm 0.72_{\text{stat}})$ GeV, which is
compatible with the input mass within one standard deviation. Also, in
the case of the extended likelihood function at NLO accuracy, the
statistical uncertainty is reduced by more than a factor of $2$. In
contrast to the Born approximation, the systematic uncertainty, as
obtained by scale variations, is also reduced at NLO and amounts to an
uncertainty of $-0.3$ GeV and $+1.8$ GeV.  Furthermore, in the
extended likelihood approach, the NLO estimator is covered by the
uncertainty of the leading-order estimator, which justifies the use of
scale variations as an estimate for the missing higher-order
corrections. As mentioned above, the parton shower does not have a
significant impact on the number of accepted events in comparison to
the fixed-order NLO calculation. However, the predicted number of
events depends approximately linearly on the value of top-quark mass
parameter~\cite{Martini:2017ydu,Kant:2014oha}.  Therefore, the
extended likelihood analysis is driven towards top-quark mass values,
which correspond to a compatible prediction of the number of events
with the pseudodata.  It is worth noting that with roughly $30000$
events the uncertainty on the extracted top-quark mass is already
dominated by systematic uncertainties.  In the above analysis, no
uncertainty for the luminosity is included.  To estimate the impact of
this additional uncertainty, we assume a relative uncertainty
$\Delta L/L\approx 2\%$~\cite{Aaboud:2016hhf}.  Repeating the extended
likelihood analysis while varying the value of $L$ by $\pm2\%$ results
in additional shifts in both the LO and the NLO estimators for the
top-quark mass of about $\pm 2$ GeV. Although, the inclusive cross
section provides important information within the MEM---pushing the
extracted mass to the value used in the simulation---we stress that the
inclusive cross section alone would lead to much larger uncertainties.
 
\section{Conclusion}\label{sec:concl}
In this article, an algorithm to calculate fully differential event
weights including NLO QCD corrections is presented. The weight is
calculated for events containing jets and not partons. We emphasise
that the weight is calculated \textit{a posteriori} for a given jet event.
This is a significant extension to existing approaches used in MC
tools which allow the generation of jet events together with the
corresponding weight but do not allow one to evaluate the weight for a
given jet event.  In contrast to previous work
\cite{Martini:2015fsa,Martini:2017ydu}, the conventional $2\to 1$
recombination is used---no change of the recombination procedure is
required.  The approach is thus directly applicable in the
experimental analysis. The method relies on properly chosen variables
to describe the jet event.  As explained in detail in the article, the
variables must not depend on the jet masses since this would spoil the
pointwise one-to-one correspondence of real and virtual
corrections. We give a concrete example for a consistent choice of
variables together with the required formulas and show how the phase space
is rewritten to allow the evaluation of the fully differential event
weight. The approach is completely general and can be applied to other
LHC processes.

The method has been used to study for the first time parton-shower
effects in the MEM@NLO.  We illustrated the impact of the parton
shower exemplarily within the context of parameter determination.  We
studied the top-quark mass extraction with the Matrix Element Method
with likelihood functions that include the fully NLO QCD corrected
event weight. Pseudodata are generated using the \textsc{POWHEG-BOX} with
subsequent showering using \textsc{Pythia8} to incorporate parton-shower
effects. We observe that a significant calibration would be required
to reconstruct the input value from the value extracted by the
MEM. This is not surprising since the parton shower leads to a
significant distortion of the distributions preferring smaller mass values.
Within the
extended likelihood approach, however, perfect agreement with the input value
for the top-quark mass of the pseudodata is found. In particular, we
show that in this approach no calibration is required and observe a
substantial reduction of the systematic uncertainties when NLO
corrections are taken into account in the calculation of the likelihood.
We expect that these findings are not specific to the example studied
here but apply also to other processes.

Actually, the approach is not restricted to
a specific process. The ability to predict event
weights at NLO accuracy for jet events enables the application of the
Matrix Element Method at NLO for events recorded by the experiments
without a modification of the recombination procedure in conventional
jet algorithms.

Finally, in
\Refs{Soper:2011cr,Soper:2012pb,Soper:2014rya,Prestel:2019neg},
information from both the leading-order matrix element as well as the
parton shower is taken into account when calculating the
likelihood. It is conceivable that combining that approach with the
method presented in this work will prove beneficial en route to an
unbiased Matrix Element Method for collider experiments.

In this article, the transfer functions are modelled as
delta-functions. Although the impact of transfer functions might be
reduced because of the better jet modelling in NLO, it is clear that a fully
realistic analysis should also include transfer functions as extracted
in the experiments.

\section*{Acknowledgments}
We would like to thank Markus Schulze for useful
discussions and careful reading of the manuscript.  This work is supported by the German Federal Ministry
for Education and Research (Grant 05H15KHCAA).

\begin{appendix}
\section{Calculational details}\label{sec:example}

In this Appendix, we provide necessary details for the implementation of the fully differential NLO weights
used in this publication.

We start the discussion with the Born phase space. Due to momentum conservation and on-shell conditions, 
each event is fully described by four variables $x_1,\ldots,x_4$. The pseudorapidity of the
top-tagged jet and the energy, the pseudorapidity and the azimuthal
angle of the (hardest) light jet are chosen.  Each set of measured values $\vec{x}=(\eta_{t},E_{j}, \eta_{j},
\phi_{j})$ specifies the
$4$-momenta of the resolved jets as functions of the squared jet masses ${J_j}^2$ and ${J_t}^2$,
\begin{eqnarray}\label{eq:eventpar21excl}
  \nn J_t&=&\left(E_{t},\;-J^{\perp}\cos{\phi_{j}},\;-J^{\perp}\sin{\phi_{j}},
    \;J^{\perp}\sinh{\eta_{t}}\right),\\
  J_{j}&=&\left(E_{j},\;J^{\perp}\cos{\phi_{j}},\;J^{\perp}\sin{\phi_{j}},
    \;J^{\perp}\sinh{\eta_{j}}\right)
\end{eqnarray}
with
\begin{displaymath}
  J^{\perp}= J^{\perp}_t= J^{\perp}_j
  ={\sqrt{{E_j}^2-{J_j}^2}\over\cosh{\eta_j}},
  \quad E_{t}=\sqrt{{J^{\perp}}^2\cosh^2{\eta_t}+{J_t}^2}
\end{displaymath}
and the Jacobian
\begin{equation}\label{eq:jacjetev}
  {\cal J}(\eta_{t},E_{j}, \eta_{j}, \phi_{j})=
  \left|\partial(J^z_t,J^x_j,J^y_j,J^z_j)\over
  \partial(\eta_{t},E_{j}, \eta_{j},
  \phi_{j})\right|={E_j\;{J^\perp_t}{J^\perp_j}
  \cosh{\eta_t}\over\cosh{\eta_j}}.
\end{equation}
From \Eq{eq:dxsbvpart}, the Born and virtual contribution of the NLO
event weight follows as
\begin{equation}\label{eq:diffxsbv}
{d^{4}\sigma^{\text{BV}}\over d{\eta_t}\;d{E_j}\;d{\eta_j}\;d{\phi_j}}
   =(2\pi)^{-2}{{J^{\perp}}^2\cosh{\eta_t}\over 2\;s
     \;E_t\;\cosh{\eta_j}} \;\left[B+V\right](x_a,x_b;J_t,J_j)
\end{equation}
with the jet momenta from \Eq{eq:eventpar21excl}.  
According to \Eq{eq:bornps}, the squared jet masses are given by ${J_t}^2={m_t}^2$
and ${J_j}^2=0$, and the momentum fractions follow as
\begin{eqnarray}\label{eq:bornxaxb}
  x_a&=&{1\over\sqrt{s}}\left(E_t+E_j
    +J^{\perp}(\sinh\eta_t+\sinh\eta_j)\right),\nn\\
  x_b&=&{1\over\sqrt{s}}\left(E_t+E_j
    -J^{\perp}(\sinh\eta_t+\sinh\eta_j)\right).
\end{eqnarray}

In the real corrections, the top quark with momentum $p_t$ and a light
parton with momentum $p_l$ are produced together with additional
radiation with momentum $p_r$:
\begin{equation}\label{eq:realreact}
p_a+p_b\rightarrow p_t+p_l+p_r\;.
\end{equation}
The phase space of the real corrections contributing to single
top-quark production in association with a light jet can thus be split
into the regions ${\cal R}_{tr}$, ${\cal R}_{lr}$, ${\cal R}_{r}$, and
$\tilde{\cal R}_{r}$.  For the clustering of the extra radiation with
the top quark, \Eq{eq:dxs21part} yields
\begin{eqnarray}\label{eq:diffxstr}
  {d^{4}\sigma^{\text{R}}_{{{\cal R}_{tr}}}\over d{\eta_t}\;d{E_j}\;d{\eta_j}\;d{\phi_j}}
  &=&(2\pi)^{-3}\frac{{J^{\perp}}^2\cosh{\eta_t}}{2s\;\cosh{\eta_j}}\;
  \int_{{{\cal R}_{tr}}}{d{J_t}^2\over E_t}\;
  d\Phi_2(J_t,\{p_t,p_r\},\{m_t,0\})\;\nn\\
  &&\quad\times\left(R(x_a,x_b;p_t,J_j,p_r)+R(x_a,x_b;p_t,p_r,J_j) \right)
\end{eqnarray}
with the jet momenta parametrisation from \Eq{eq:eventpar21excl}. 
Because of the on-shell conditions, the squared jet masses follow 
as ${J_t}^2={M_{tr}}^2=(p_t+p_r)^2$ and ${J_j}^2=0$. The parton momentum fractions 
are again given by \Eq{eq:bornxaxb}. 
Note that \Eq{eq:diffxstr} already takes into account that either of the massless quarks can be clustered 
together with the top quark, while the other one constitutes the light jet.

In the case where the light jet is obtained from the clustering of the
light quark and the additional radiation, \Eq{eq:dxs21part} yields
\begin{eqnarray}\label{eq:diffxslr}
  {d^{4}\sigma^{\text{R}}_{{{\cal R}_{lr}}}
    \over  d{\eta_t}\;d{E_j}\;d{\eta_j}\;d{\phi_j}}
  &=&(2\pi)^{-3}\frac{\cosh{\eta_t}}{2s\;\cosh{\eta_j}}\;
  \int_{{{\cal R}_{lr}}}{d{J_j}^2}\;{{J^{\perp}}^2\over E_t}\;
  d\Phi_2(J_j,\{p_l,p_r\},\{0,0\})\;  R(x_a,x_b;J_t,p_l,p_r) \nn\\
\end{eqnarray}
with the jet momenta from \Eq{eq:eventpar21excl}. The on-shell conditions result in ${J_t}^2={m_t}^2$
and ${J_j}^2={M_{lr}}^2=(p_l+p_r)^2$, while the parton momentum fractions are again
given by \Eq{eq:bornxaxb}.  

In the regions of the phase space where
the extra radiation is associated with the beam, the jet momenta have
to be parametrised by the variables $(\eta_{t},E_{j}, \eta_{j},
\phi_{j})$ and the $3$-momentum of the extra radiation in order to
ensure momentum conservation,
\begin{eqnarray}\label{eq:eventparincl}
\nn  J_t&=&\left(E_{t},\;-J^{\perp}_t\cos{\phi_{t}},\;
    -J^{\perp}_t\sin{\phi_{t}},\;J^{\perp}_t\sinh{\eta_{t}}\right)\;,\\
 J_{j}&=&\left(E_{j},\;J^{\perp}_j\cos{\phi_{j}},\;J^{\perp}_j
    \sin{\phi_{j}},\;J^{\perp}_j\sinh{\eta_{j}}\right)
\end{eqnarray}
with
\begin{eqnarray*}
  &J^{\perp}_j=\displaystyle{E_{j}\over\cosh{\eta_j}}\;,\quad J^{\perp}_t
  =\sqrt{\left(J^{\perp}_j\cos{\phi_{j}}+p^{x}_{r}\right)^2
    +\left(J^{\perp}_j\sin{\phi_{j}}+p^{y}_{r}\right)^2}\;,&\\
  &\tan\phi_t=\displaystyle{J^{\perp}_j\sin{\phi_{j}}+p^{y}_{r}
    \over J^{\perp}_j\cos{\phi_{j}}+p^{x}_{r}}\;,
  \quad
  E_{t}=\sqrt{{J^{\perp}_t}^2\cosh^2{\eta_t}+m^2_t}\;.
\end{eqnarray*}
Note that the jet momenta defined in \Eq{eq:eventparincl} fulfill the on-shell conditions in \Eq{eq:realise} by construction: ${J_t}^2={m_t}^2$ and ${J_j}^2=0$.
Using this parametrisation and \Eq{eq:dxsinipart} together with
\Eq{eq:jacjetev} yields
\begin{eqnarray}\label{eq:diffxsr}
  {d^{4}\sigma^{\text{R}}_{{{\cal R}_{r}}}\over
    d{\eta_t}\;d{E_j}\;d{\eta_j}\;d{\phi_j}}
  &=&(2\pi)^{-5}{{J^{\perp}_j}\cosh{\eta_t}\over4s\;\cosh{\eta_j}}\;
  \int_{{{\cal R}_{r}}}{d^3p_{r}}\;{{J^{\perp}_t}\over E_t\;|\vec{p}_{r}|}\nn\\
  &&\quad\times\left(R(x_a,x_b;J_t,J_j,p_r)+R(x_a,x_b;J_t,p_r,J_j)\right)
\end{eqnarray}
with the parton momentum fractions given by
\begin{eqnarray}\label{eq:realxaxb}
  x_a&=&{1\over\sqrt{s}}\left(E_t+E_j+|\vec{p}_{r}|
    +J^{\perp}_t\sinh\eta_t+J^{\perp}_j\sinh\eta_j+p^z_r\right)\;,\nn\\ 
  x_b&=&{1\over\sqrt{s}}\left(E_t+E_j+|\vec{p}_{r}|
    -J^{\perp}_t\sinh\eta_t-J^{\perp}_j\sinh\eta_j-p^z_r\right)\;.
\end{eqnarray}
Again, either of the massless quarks can escape detection, while the other one constitutes the light jet.

The region $\tilde {\cal R}_{r}$, corresponding to the extra radiation
being resolved as an additional but softer light jet, contributes to
the event weight as (cf. \Eq{eq:dxsaddjpart})
\begin{eqnarray}\label{eq:diffxsj}
 {d^{4}\sigma^{\text{R}}_{{\tilde{\cal R}_{r}}}\over 
       d{\eta_t}\;d{E_j}\;d{\eta_j}\;d{\phi_j}}
  &=&(2\pi)^{-5}{{J^{\perp}_j}\cosh{\eta_t}\over4s\;\cosh{\eta_j}}\;\int_{{\tilde{\cal R}_{r}}}{d^3J_{r}}\;{{J^{\perp}_t}\over E_t\;|\vec{J}_{r}|}\;\Theta\left(J^{\perp}_j-J^{\perp}_r\right)\nn\\
&&\quad\times\left(R(x_a,x_b;J_t,J_j,J_r)+R(x_a,x_b;J_t,J_r,J_j)\right)
\end{eqnarray}
with the parametrisation of the jet momenta and parton momentum fractions given in \Eq{eq:eventparincl} and \Eq{eq:realxaxb} with $p_r\rightarrow J_r$.
According to \Eq{eq:masterfac}, the weight including NLO corrections
for a jet event $t,j$ defined by $\eta_{t}$, $E_{j}$, $\eta_{j}$, and
$\phi_{j}$ is given by the sum of the Born and virtual contribution
(see \Eq{eq:diffxsbv}) and the real corrections from the regions ${\cal
  R}_{tr}$ (see \Eq{eq:diffxstr}), ${\cal R}_{lr}$ (see
\Eq{eq:diffxslr}), ${\cal R}_{r}$ (see \Eq{eq:diffxsr}), and
$\tilde{\cal R}_{r}$ (see \Eq{eq:diffxsj}).  We stress that a $2\to 1$
recombination is used where the momentum of the resulting jet is
defined as the $4$-momentum sum of the recombined partons. 

\section{Validation}\label{sec:valid}
In this Appendix, we show a short validation of the presented method. To this end, we generated
approximately $40000$ unweighted events at NLO accuracy to compute differential distributions, which 
are compared to an ordinary Monte-Carlo integration. The input parameters for the simulation
are the same as described in section~\ref{sec:example2}, which for brevity we do not repeat here.

\begin{figure*}[htbp]
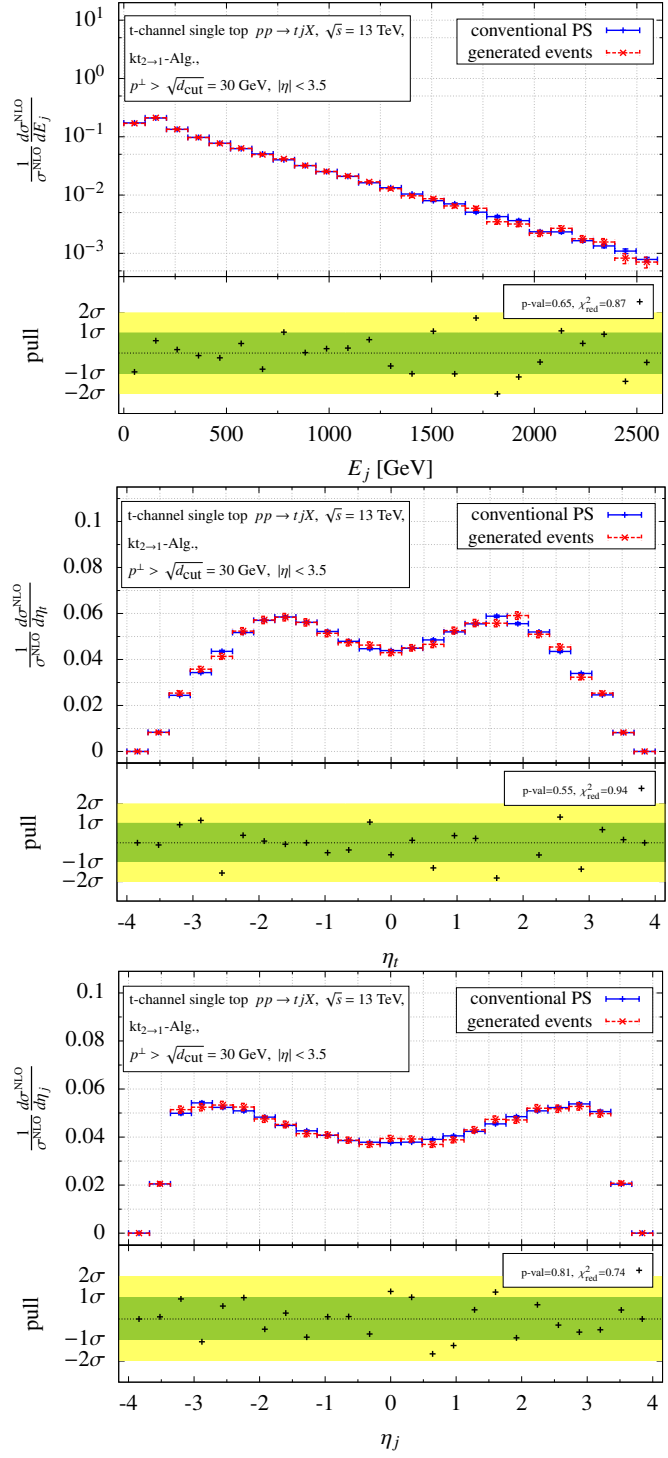

  \begin{center}
    \leavevmode
     \includegraphics[height=0.305\textheight]{{{%
          sgttKT3-2ycut30comptwoincE2-127evts-crop}}}
     \includegraphics[height=0.305\textheight]{{{%
          sgttKT3-2ycut30comptwoincETA1-127evts-crop}}}
     \includegraphics[height=0.305\textheight]{{{%
          sgttKT3-2ycut30comptwoincETA2-127evts-crop}}}
    \caption{Energy and pseudorapidity distributions for $t$-channel 
      single top-quark production calculated at NLO accuracy using a
      conventional parton-level Monte Carlo with $2\to1$ jet clustering (solid
      blue) compared to histograms filled with generated NLO events
      (dashed red). At the bottom of the plots, the pull distributions
      together with the p-value and the reduced $\chi^2$ of the
      histogram comparisons are shown.}
    \label{fig:partonMCvsgenev_sgtt21}
  \end{center}
\end{figure*}

The red dashed histograms in \Fig{fig:partonMCvsgenev_sgtt21} show the $E_j$, $n_j$, and 
$\eta_t$ distributions obtained from the unweighted NLO event sample.  These distributions
are compared with the results of independent calculations using a
parton level Monte-Carlo integration (blue solid histogram in
\Fig{fig:partonMCvsgenev_sgtt21}).  In the lower part of the plots, the
respective pull distributions, p-values, and reduced $\chi^2$ of the
comparison of the two histograms as described in
\Ref{Gagunashvili:2007zz} and implemented in
\Refs{Brun:1997pa,Antcheva:2009zz,Moneta:2008zza} are given.  The
results in \Fig{fig:partonMCvsgenev_sgtt21} show that the unweighted
events are indeed distributed according to the cross section
calculated at NLO accuracy. Within the statistical uncertainties, the
results from the parton-level Monte Carlo are in perfect agreement
with the results obtained from the sample of unweighted events.
\end{appendix}

\providecommand{\href}[2]{#2}\begingroup\raggedright\endgroup
\end{document}